\documentclass[12pt]{article}

\usepackage{amsmath}
\usepackage{amssymb}
\usepackage[dvips]{graphicx}
\usepackage{cite}

\makeatletter
 
  \@addtoreset{equation}{section}
 \makeatother

\newcommand {\n}{\nonumber \\}
\newcommand {\tr}{\mbox{tr}}
\newcommand {\Tr}{\mbox{Tr}}

\begin{document}
\setlength{\oddsidemargin}{0cm}
\setlength{\baselineskip}{7mm}

\begin{titlepage}
\begin{normalsize}
\begin{flushright}
\begin{tabular}{l}
October 2011
\end{tabular}
\end{flushright}
  \end{normalsize}

~~\\

\vspace*{0cm}
    \begin{Large}
       \begin{center}
         {Supersymmetry and DLCQ Limit 

of 

Lie 3-algebra Model of M-theory}
       \end{center}
    \end{Large}
\vspace{1cm}

\begin{center}
           Matsuo S{\sc ato}\footnote
           {
e-mail address : msato@cc.hirosaki-u.ac.jp}\\
      \vspace{1cm}
       
         {\it Department of Natural Science, Faculty of Education, Hirosaki University\\ 
 Bunkyo-cho 1, Hirosaki, Aomori 036-8560, Japan}

\end{center}

\hspace{5cm}

\begin{abstract}
\noindent
In arXiv:1003.4694, we proposed two models of M-theory, Hermitian 3-algebra model and Lie 3-algebra model. In this paper, we study the Lie 3-algebra model with a Lorentzian Lie 3-algebra. This model is ghost-free despite the Lorentzian 3-algebra. We show that our model satisfies two criteria as a model of M-theory. First, we show that the model possesses $\mathcal{N}=1$ supersymmetry in eleven dimensions. Second, we show the model reduces to BFSS matrix theory with finite size matrices in a DLCQ limit.
\end{abstract}

\vfill
\end{titlepage}
\vfil\eject

\setcounter{footnote}{0}

\section{Introduction}
\setcounter{equation}{0}

BFSS matrix theory \cite{BFSS} is one of the strong candidates of non-perturbative definition of superstring theory. It is conjectured to describe infinite momentum frame (IMF) limit of M-theory and many evidences were found. Because only D0-branes in type IIA superstring theory survive in this limit, BFSS matrix theory is defined by the one-dimensional maximally supersymmetric Yang-Mills theory. Since the theory is a gauge theory, a matrix representation is allowed and dynamics of a many-body system can be described by using diagonal blocks of matrices. However, it seems impossible to derive full dynamics of M-theory from BFSS matrix theory because it treats D0-branes as fundamental degrees of freedom. For example, we do not know the manner to describe longitudinal momentum transfer of D0-branes. Therefore, we need a matrix model that treats membranes as fundamental degrees of freedom in order to derive full dynamics of M-theory.

IIB matrix model \cite{IKKT} is also one of the strong candidates of non-perturbative definition of superstring theory. It starts with the Green-Schwartz type IIB superstring action in order to treat strings themselves as fundamental degrees of freedom. If we fix the $\kappa$ symmetry to Schild gauge $\theta_1=\theta_2$, the action reduces to that of the zero-dimensional maximal supersymmetric Yang-Mills theory with area preserving diffeomorphism (APD) symmetry. Since the resultant action is a gauge theory, it describes dynamics of many-body systems. IIB matrix model is defined by replacing the APD algebra with $u(N)$ Lie algebra in the action. 

In paper \cite{MModel}, we obtained matrix models of M-theory in an analogous way to obtain IIB matrix model. We started with the Green-Schwartz supermembrane action in order to obtain matrix models of M-theory that treat membranes themselves as fundamental degrees of freedom. We showed, by using an approximation, that the action reduces to that of a zero-dimensional gauge theory with volume preserving diffeomorphism (VPD) symmetry \cite{Nambu, Yoneya} if we fix the $\kappa$ symmetry of the action to a semi-light-cone gauge, $\Gamma_{012} \Psi = \Psi$. We proposed two 3-algebra models of M-theory which are defined by replacing VPD algebra with finite-dimensional 3-algebras in the action. Because the 3-algebra models are gauge theories, they are expected to describe dynamics of many-body systems as in the other matrix models.

One of the two models is based on Hermitian 3-algebra \cite{ABJM, N=6BL, LieOrigin, CherkisDotsenkoSaeman} (Hermitian 3-algebra model), whereas the another is based on Lie 3-algebra \cite{Filippov, Kamiya, OkuboKamiya, BLG1, Gustavsson, BLG2, Nogo1, Nogo2, Nogo3, Lorentz1, Lorentz2, Lorentz3, Ghost-Free} (Lie 3-algebra model). The Hermitian 3-algebra model with $u(N) \oplus u(N)$ symmetry was shown to reduce to BFSS matrix theory with finite size matrices when a DLCQ limit is taken in \cite{MModel}. A supersymmetric deformation of the Lie 3-algebra model with the $\mathcal{A}_4$ algebra was investigated by adding mass and flux terms in \cite{DeBellisSaemannSzabo}.

In this paper, we study the Lie 3-algebra model with a Lorentzian Lie 3-algebra. We show that this model satisfies two criterion as a model of M-theory. In section two, we show that the model possesses $\mathcal{N}=1$ supersymmetry in eleven dimensions. In section three, we show the model reduces to BFSS matrix theory with finite size matrices in a DLCQ limit as it should do: it is generally shown that M-theory reduces to such BFSS matrix theory in a DLCQ limit \cite{Susskind, Sen, Seiberg, Polchinski}.

\vspace{1cm}

\section{$\mathcal{N}=1$ Supersymmetry Algebra in Eleven Dimensions}
\setcounter{equation}{0}
In \cite{MModel}, we proposed the Lie 3-algebra model of M-theory, whose action is given by
\begin{eqnarray}
S_{0}&=&
\Bigl<
-\frac{1}{12}[X^I, X^J, X^K]^2 
-\frac{1}{2}(A_{\mu a b} [T^a, T^b, X^{I}])^2 \n
&& \quad 
-\frac{1}{3} E^{\mu \nu \lambda} 
A_{\mu a b} A_{\nu c d} A_{\lambda e f} 
[T^a, T^c, T^d][T^b, T^e, T^f] \n
&& \quad 
-\frac{i}{2}\bar{\Psi} \Gamma^{\mu} A_{\mu a b} [T^a, T^b, \Psi] 
+\frac{i}{4}\bar{\Psi}\Gamma_{IJ}[X^I, X^J, \Psi] 
\Bigr>. \label{3algebraCovariantAction}
\end{eqnarray}
The fields are spanned by Lie 3-algebra $T^a$ as $X^I=X^I_a T^a$, $\Psi = \Psi_a T^a$ and $A^{\mu}= A^{\mu}_{ab} T^a \otimes T^b$, where $I=3, \cdots, 10$ and $\mu=0,1,2$. $<>$ represents a metric for the 3-algebra. $\Psi$ is a Majorana spinor of SO(1,10) that satisfies $\Gamma_{012} \Psi =\Psi$. $ E^{\mu \nu \lambda} $ is a Levi-Civita symbol in three-dimensions. In this section, we will show that this action possesses $\mathcal{N}=1$ supersymmetry in eleven-dimensions.

The action is invariant under 16 dynamical supersymmetry transformations,
\begin{eqnarray}
&&\delta X^I =i \bar{\epsilon} \Gamma^I \Psi \n
&&\delta A_{\mu a b}[T^a, T^b,\quad]=i\bar{\epsilon} \Gamma_{\mu} \Gamma_I [X^I, \Psi,\quad] \n
&&\delta \Psi = -A_{\mu a b}[T^a, T^b, X^I] \Gamma^{\mu} \Gamma_I \epsilon -\frac{1}{6}[X^I, X^J, X^K] \Gamma_{IJK} \epsilon,  \label{dynamicalSUSYtrans}
\end{eqnarray}
where $\Gamma_{012}\epsilon=-\epsilon$.
These supersymmetries close into gauge transformations on-shell, 
\begin{eqnarray}
&& [\delta_1, \delta_2]X^I=\Lambda_{cd}[T^c, T^d, X^I] \n
&& [\delta_1, \delta_2] A_{\mu a b}[T^a, T^b,\quad]
= \Lambda_{ab}[T^a, T^b, A_{\mu c d}[T^c, T^d,\quad]]-
A_{\mu a b}[T^a, T^b, \Lambda_{c d}[T^c, T^d,\quad]]+
2i\bar{\epsilon}_2 \Gamma^{\nu}\epsilon_1 O^A_{\mu \nu} \n
&& [\delta_1, \delta_2]\Psi = \Lambda_{cd}[T^c, T^d, \Psi] 
+ (i\bar{\epsilon}_2 \Gamma^{\mu} \epsilon_1 \Gamma_{\mu} -\frac{i}{4} \bar{\epsilon}_2 \Gamma^{KL} \epsilon_1 \Gamma_{KL}) O^{\Psi}, \label{dynamicalalg}
\end{eqnarray}
where gauge parameters are given by $\Lambda_{ab}=2i \bar{\epsilon}_2 \Gamma^{\mu} \epsilon_1 A_{\mu a b}-i \bar{\epsilon}_2 \Gamma_{JK} \epsilon_1 X^J_a X^K_b$. $O^A_{\mu\nu}=0$ and $O^{\Psi}=0$ are equations of motions of $A_{\mu\nu}$ and $\Psi$, respectively, where
\begin{eqnarray}
O^A_{\mu\nu}&=& A_{\mu a b}[T^a, T^b, A_{\nu c d}[T^c, T^d,\quad]]-
A_{\nu a b}[T^a, T^b, A_{\mu c d}[T^c, T^d,\quad]] \n
&&+E_{\mu \nu \lambda}(-[X^I, A^{\lambda}_{a b} [T^a, T^b, X_I],\quad]
+\frac{i}{2}[\bar{\Psi}, \Gamma^{\lambda} \Psi,\quad]) \n
O^{\Psi}&=&-\Gamma^{\mu}A_{\mu a b}[T^a, T^b, \Psi] +\frac{1}{2}\Gamma_{IJ}[X^I, X^J, \Psi]. 
\end{eqnarray}
(\ref{dynamicalalg}) implies that a commutation relation between the dynamical supersymmetry transformations is 
\begin{equation}
\delta_2 \delta_1 - \delta_1 \delta_2 =0,
\end{equation}
up to the equations of motions and the gauge transformations.

Lie 3-algebra with an invariant metric is classified into four-dimensional Euclidean $\mathcal{A}_4$ algebra and Lie 3-algebras with indefinite metrics in \cite{Nogo1, Nogo2, Nogo3, kac, Class}. We do not choose $\mathcal{A}_4$ algebra because its degrees of freedom are just four. We need an algebra with arbitrary dimensions N, which is taken to infinity to define M-theory.  Here we choose the most simple indefinite metric Lie 3-algebra, so called Lorentzian Lie 3-algebra associated with $u(N)$ Lie algebra,
\begin{eqnarray}
&&[T^{-1}, T^a, T^b]=0  \n
&&[T^{0}, T^i, T^j]=[T^i, T^j]=f^{ij}_{\quad k} T^k \n
&&[T^{i}, T^{j}, T^{k}]=f^{ijk} T^{-1},
\end{eqnarray}
where $a=-1,0,i$ ($i=1, \cdots, N^2$). $T^i$ are generators of $u(N)$. A metric is defined by a symmetric bilinear form,
\begin{eqnarray}
<T^{-1}, T^0>&=&-1 \label{indefinitemetric} \\
<T^i, T^j>&=&h^{ij}, 
\end{eqnarray}
and the other components are 0. The action is decomposed as
\begin{eqnarray}
S=\Tr(
-\frac{1}{4}(x_0^K)^2[x^I, x^J]^2 
+\frac{1}{2}(x_0^I[x_I, x^J])^2
-\frac{1}{2}(x_0^I b_{\mu} + [a_{\mu}, x^I])^2
-\frac{1}{2}E^{\mu \nu \lambda} b_{\mu}[a_{\nu}, a_{\lambda}] \n
 +i \bar{\psi}_0 \Gamma^{\mu} b_{\mu} \psi
-\frac{i}{2} \bar{\psi} \Gamma^{\mu}[a_{\mu}, \psi]
+\frac{i}{2} x_0^I \bar{\psi} \Gamma_{IJ}[x^J, \psi]
-\frac{i}{2}\bar{\psi}_0 \Gamma_{IJ}[x^I, x^J]\psi),
\label{ExplicitAction}
\end{eqnarray}
where we have renamed 
$X_0^I \to x_0^I$,
$X^I_i T^i \to x^I$,
$\Psi_0 \to \psi_0$,
$\Psi_i T^i \to \psi$,
$2A_{\mu 0 i} T^i \to a_{\mu}$, and
$A_{\mu i j}[T^i, T^j] \to b_{\mu}$. 
In this action, $T^{-1}$ mode; $X^I_{-1}$, $\Psi_{-1}$ or $A^{\mu}_{-1 a}$ does not appear, that is they are unphysical modes. Therefore, the indefinite part of the metric (\ref{indefinitemetric}) does not exist in the action and our model is ghost-free like a model in \cite{bosonicM}.  This action can be obtained by a dimensional reduction of the three-dimensional $\mathcal{N}=8$ BLG model \cite{BLG1, Gustavsson, BLG2} with the same 3-algebra. The BLG model possesses a ghost mode because of its kinetic terms with indefinite signature. On the other hand, our model does not possess a kinetic term because it is defined as a zero-dimensional field theory like IIB matrix model \cite{IKKT}. 

This action is invariant under the translation 
\begin{equation}
\delta x^I = \eta^I, \qquad \delta a^{\mu} = \eta^{\mu},
\end{equation}
where $\eta^I$ and $\eta^{\mu}$ belong to $u(1)$. This implies that eigen values of $x^I$ and $a^{\mu}$ represent an eleven-dimensional space-time.  

The action is also invariant under 16 kinematical supersymmetry transformations
\begin{equation}
\tilde{\delta} \psi = \tilde{\epsilon},
\end{equation}
and the other fields are not transformed. $\tilde{\epsilon}$ belong to $u(1)$ and satisfy $\Gamma_{012}\tilde{\epsilon}=\tilde{\epsilon}$. $\tilde{\epsilon}$ and $\epsilon$ should come from 16 components of 32 $\mathcal{N}=1$ supersymmetry parameters in eleven dimensions, corresponding to eigen values $\pm$1 of $\Gamma_{012}$, respectively, as in the case of the semi-light-cone supermembrane. Its target-space $\mathcal{N}=1$ supersymmetry consists of remaining 16 target-space supersymmetries and transmuted 16 $\kappa$-symmetries in the semi-light-cone gauge, $\Gamma_{012} \Psi = \Psi$ \cite{MModel, deWHN, MSUSY}.

A commutation relation between the kinematical supersymmetry transformations is given by
\begin{equation}
\tilde{\delta}_2 \tilde{\delta}_1 - \tilde{\delta}_1 \tilde{\delta}_2=0.
\end{equation}
The 16 dynamical supersymmetry transformations (\ref{dynamicalSUSYtrans}) are decomposed as
\begin{eqnarray}
&&\delta x^I =i \bar{\epsilon} \Gamma^I \psi \n
&&\delta x_0^I =i \bar{\epsilon} \Gamma^I \psi_0 \n
&&\delta x_{-1}^I =i \bar{\epsilon} \Gamma^I \psi_{-1} \n
\n
&&\delta \psi=-(b_{\mu} x^I_0 +[a_{\mu}, x^I])\Gamma^{\mu}\Gamma_I \epsilon
-\frac{1}{2}x_0^I[x^J, x^K] \Gamma_{IJK} \epsilon \n
&&\delta \psi_0 =0 \n
&&\delta\psi_{-1}=-\Tr(b_{\mu} x^I)\Gamma^{\mu}\Gamma_I \epsilon 
-\frac{1}{6}\Tr([x^I, x^J] x^K)\Gamma_{IJK} \epsilon \n
\n
&&\delta a_{\mu} =i \bar{\epsilon} \Gamma_{\mu} \Gamma_I (x_0^I \psi -\psi_0 x^I) \n
&&\delta b_{\mu} =i \bar{\epsilon} \Gamma_{\mu} \Gamma_I [x^I, \psi] \n
&&\delta A_{\mu -1 i} =i \bar{\epsilon} \Gamma_{\mu} \Gamma_{I} \frac{1}{2}(x_{-1}^I \psi_i -\psi_{-1} x_i^I) \n
&&\delta A_{\mu -1 0} =i \bar{\epsilon} \Gamma_{\mu} \Gamma_{I} \frac{1}{2}(x_{-1}^I \psi_0 -\psi_{-1} x_0^I),
\end{eqnarray}
and thus a commutator of dynamical supersymmetry transformations and kinematical ones acts as
\begin{eqnarray}
&&(\tilde{\delta}_2 \delta_1 - \delta_1 \tilde{\delta}_2)x^I
=i \bar{\epsilon}_1 \Gamma^I \tilde{\epsilon}_2 
\equiv \eta^I \n
&&(\tilde{\delta}_2 \delta_1 - \delta_1 \tilde{\delta}_2)a^{\mu}
=i \bar{\epsilon}_1 \Gamma^{\mu} \Gamma_I x_0^I \tilde{\epsilon}_2 
\equiv \eta^{\mu} \n
&&(\tilde{\delta}_2 \delta_1 - \delta_1 \tilde{\delta}_2) A_{-1 i}^{\mu}T^i
= \frac{1}{2} i \bar{\epsilon}_1 \Gamma^{\mu} \Gamma_I x^I_{-1} \tilde{\epsilon}_2,
\end{eqnarray}
where the commutator that acts on the other fields vanishes. Thus, the commutation relation for physical modes is given by  
\begin{equation}
\tilde{\delta}_2 \delta_1 - \delta_1 \tilde{\delta}_2 = \delta_{\eta},
\end{equation}\
where $\delta_{\eta}$ is a translation.


If we change a basis of the supersymmetry transformations as
\begin{eqnarray}
&&\delta'= \delta + \tilde{\delta} \n
&&\tilde{\delta}' = i(\delta- \tilde{\delta}),
\end{eqnarray}
we obtain 
\begin{eqnarray}
&&\delta'_2 \delta'_1 - \delta'_1 \delta'_2 = \delta_{\eta} \n
&&\tilde{\delta}'_2 \tilde{\delta}'_1 - \tilde{\delta}'_1 \tilde{\delta}'_2 = \delta_{\eta} \n
&&\tilde{\delta}'_2 \delta'_1 - \delta'_1 \tilde{\delta}'_2 = 0. \label{preN=1}
\end{eqnarray}
These 32 supersymmetry transformations are summarised as $\Delta=(\delta', \tilde{\delta}')$ and (\ref{preN=1}) implies the $\mathcal{N}=1$ supersymmetry algebra in eleven dimensions,
\begin{equation}
\Delta_2 \Delta_1 -\Delta_1 \Delta_2 = \delta_{\eta}.
\end{equation}

\section{DLCQ limit}
\setcounter{equation}{0}
In this section, we will take a DLCQ limit of our model and obtain BFSS matrix theory with finite size matrices as desired. 

First, we separate the auxiliary fields $b^{\mu}$ from $A^{\mu}$ and define $X^{\mu}$ by 
\begin{equation}
A^{\mu}=X^{\mu}+b^{\mu}.
\end{equation}
We identify space-time coordinate matrices by redefining matrices as follows. By rescaling the eight matrices as
\begin{eqnarray}
&&X^I=\frac{1}{T}X'^I \n
&&X^{\mu}= X'^{\mu},
\label{StartLimit}
\end{eqnarray}
we adjust the scale of $X^I$ to that of $X^{\mu}$. $T$ is a real parameter. Next, we redefine fields so as to keep the scale of nine matrices:
\begin{eqnarray}
&&X'^p=X''^p \n
&&X'^i=X''^i \n
&&X'^0=\frac{1}{T}X''^0 \n
&&X'^{10}=\frac{1}{T}X''^{10} 
\end{eqnarray}
where $p=1,2$ and $i=3, \cdots, 9$. 
We also redefine the auxiliary fields as 
\begin{equation}
b^{\mu}=\frac{1}{T^2} b'''^{\mu}.
\end{equation}

DLCQ limit of M-theory consists of a light-cone compactification, $x^- \approx x^- + 2 \pi R$, where $x^{\pm}=\frac{1}{\sqrt{2}}(x^{10} \pm x^0)$, and Lorentz boost in $x^{10}$ direction with an infinite momentum. We define light-cone coordinates as
\begin{eqnarray}
&&X''^{0} = \frac{1}{\sqrt{2}}(X^{+}-X^{-}) \n
&&X''^{10} = \frac{1}{\sqrt{2}}(X^{+}+X^{-}) 
\end{eqnarray}
We treat $b^{''' \mu}$ as scalars.  
A matrix compactification \cite{Taylor} on a circle with a radius R
imposes following conditions on $X^{-}$ and the other matrices $Y$, which represent $X^+$, $X^{''p}$, $X^{''i}$, $b^{''' \mu}$, and $\Psi$:
\begin{eqnarray}
&&X^- -(2 \pi R) \bold{1} = U^{\dagger} X^- U \n
&&Y = U^{\dagger} Y U,
\label{periodic}
\end{eqnarray}
where $U$ is a unitary matrix. After the compactification, we cannot redefine fields freely.
A solution to (\ref{periodic}) is given by $X^{-} = \bar{X}^{-} + \tilde{X}^{-}$, $Y=\tilde{Y}$ and 
\begin{equation}
U=\tilde{U} \otimes \bold{1}_{\mbox{Lorentzian}},
\end{equation}
where U(N) part is given by,
\begin{equation}
\tilde{U}=
\left(
\begin{array}{cccc}
0&1&&0 \\
&\ddots&\ddots& \\
&&&1 \\
0&&&0
\end{array}
\right)
\otimes
\bold{1}_{n \times n}.
\end{equation}
A background $\bar{X}^{-}$ is
\begin{equation}
\bar{X}^- = - T^3 \bar{x}_0^- T^0 -(2 \pi R) \mbox{diag}(\cdots, s-1, s, s+1, \cdots) \otimes \bold{1}_{n \times n},
\label{bg}
\end{equation}
and a fluctuation $\tilde{x}$ that represents $u(N)$ parts of $\tilde{X}^{-}$ and $\tilde{Y}$ is
\begin{equation} 
\left(
\begin{array}{cccc}
\tilde{x}(0)&\tilde{x}(1)& \cdots & \\
\tilde{x}(-1)&\ddots&\ddots & \\
\vdots&\ddots&&\tilde{x}(1) \\
&&\tilde{x}(-1)&\tilde{x}(0)
\end{array}
\right). \label{fluctuation}
\end{equation}
Each $\tilde{x}(s)$ is a $n \times n$ matrix, where $s$ is an integer. That is, the (s, t)-th block is given by $\tilde{x}_{s,t} = \tilde{x}(s-t)$.

We make a Fourier transformation,
\begin{equation}
\tilde{x}(s)= \frac{1}{2 \pi \tilde{R}} \int^{2 \pi \tilde{R}}_0 d\tau x(\tau) e^{i s \frac{\tau}{\tilde{R}}},  \label{Fourier}
\end{equation}
where $x(\tau)$ is a $n \times n$ matrix in one-dimension and $R\tilde{R}=2\pi$. From (\ref{bg}), (\ref{fluctuation}) and (\ref{Fourier}), the following identities hold:
\begin{eqnarray}
&&
\sum_t \tilde{x}_{s,t} \tilde{x'}_{t, u}
=
\frac{1}{2 \pi \tilde{R}} \int^{2 \pi \tilde{R}}_0 d\tau \,
x(\tau) x'(\tau) e^{i(s-u) \frac{\tau}{\tilde{R}}} \n
&&
\tr(\sum_{s, t} \tilde{x}_{s,t} \tilde{x'}_{t,s})
=
V\frac{1}{2 \pi \tilde{R}} \int^{2 \pi \tilde{R}}_0 d\tau \, \tr(x(\tau)x'(\tau)) \n
&&
[\bar{x}^-, \tilde{x}]_{s, t} 
=
\frac{1}{2 \pi \tilde{R}} \int^{2 \pi \tilde{R}}_0 d\tau \,  \partial_{\tau} x(\tau) 
e^{i(s-t) \frac{\tau}{\tilde{R}}}, \label{id}
\end{eqnarray}
where $\tr$ is a trace over $n \times n$ matrices and $V=\sum_s 1$. We will use these identities later.

Next, let us boost the system in $x^{10}$ direction: 
\begin{eqnarray}
&&\tilde{X}^+ = \frac{1}{T} \tilde{X}'''^+ \n
&&\tilde{X}^- = T \tilde{X}'''^- \n
&&\tilde{X}''^p = \tilde{X}'''^p \n
&&\tilde{X}''^i = \tilde{X}'''^i.
\end{eqnarray}
IMF limit is achieved when $T \to \infty$.  
The second equation implies that $X^- = -T^3 \bar{x}_0^- T^0 + T X'''^-$,
where
$X'''^- = \bar{X}'''^- + \tilde{X}'''^- $
and $ \bar{X}'''^- = -(2 \pi R') \mbox{diag} (\cdots, s-1, s, s+1, \cdots) \otimes \bold{1}_{n \times n}$. $R' = \frac{1}{T} R$ goes to zero when $T \to \infty$. 
To keep supersymmetry, the fermionic fields need to behave as
\begin{equation}
\Psi = \frac{1}{T} \Psi'''.
\end{equation}

To summarize, relations between the original fields and the fixed fields when $T \to \infty$ are 
\begin{eqnarray}
&&a^0= \frac{1}{\sqrt{2}}(\frac{1}{T^2}x'''^+ - x'''^-) \n
&&a^p = x'''^p \n
&&x^i = \frac{1}{T} x'''^i \n
&&x^{10}=\frac{1}{\sqrt{2}}(\frac{1}{T^3}x'''^+ + \frac{1}{T}x'''^-) \n
&&x^i_0 = \frac{1}{T} x'''^i_0 \n
&&x^{10}_0= \frac{1}{\sqrt{2}}(\frac{1}{T^3}x'''^+_0 + \frac{1}{T} x'''^-_0) - \frac{1}{\sqrt{2}} T \bar{x}^-_0 \n
&&b^{\mu} = \frac{1}{T^2} b'''^{\mu} \n
&& \psi = \frac{1}{T} \psi''' \n
&&\psi_0 = \frac{1}{T} \psi'''_0.
\label{SummaryLimit}
\end{eqnarray}
By using these relations, equations of motion of the auxiliary fields $b^{\mu}$,\begin{equation}
b^{\mu}=\frac{1}{(x_0^I)^2}(-x_0^I[a^{\mu}, x_I] -\frac{1}{2}E^{\mu \nu \lambda}[a_{\nu}, a_{\lambda}] +i \bar{\psi}_0 \Gamma^{\mu} \psi)
\end{equation}
are rewritten as
\begin{eqnarray}
&&b'''^0 = -\frac{2}{(\bar{x}^-_0)^2}[x'''^1, x'''^2] + O(\frac{1}{T}) \n
&&b'''^1 = (-\frac{\sqrt{2}}{(\bar{x}^-_0)^2}[x'''^2, x'''^-] +\frac{1}{\bar{x}^-_0}[x'''^1, x'''^-])+ O(\frac{1}{T}) \n
&&b'''^2 = (\frac{\sqrt{2}}{(\bar{x}^-_0)^2}[x'''^1, x'''^-] +\frac{1}{\bar{x}^-_0}[x'''^2, x'''^-])+ O(\frac{1}{T}).
\end{eqnarray}
If we substitute them and (\ref{SummaryLimit}) to the action (\ref{ExplicitAction}), we obtain
\begin{eqnarray}
&& \! \! \! \! S=\frac{1}{T^2} \Tr (\frac{1}{2(\bar{x}^-_0)^2}[x'''^-, x'''^p]^2+ \frac{1}{4}[x'''^-, x'''^i]^2
-\frac{1}{2(\bar{x}^-_0)^2}[x'''^p, x'''^q]^2
-\frac{1}{2}[x'''^p, x'''^i]^2
-\frac{(\bar{x}^-_0)^2}{8}[x'''^i, x'''^j]^2 \n
&& \qquad \qquad -\frac{i}{2\sqrt{2}} \bar{\psi}''' \Gamma^0 [x'''^-, \psi''']
-\frac{i}{2} \bar{\psi}''' \Gamma^p [x'''_p, \psi''']
-\frac{i}{2\sqrt{2}} \bar{x}^-_0 \bar{\psi}''' \Gamma_{10 i} [x'''^i, \psi'''])
+O(\frac{1}{T^3}).
\end{eqnarray}
Therefore, the action reduces to
\begin{eqnarray}
&& \! \! \! \! \hat{S}= \frac{1}{T^2} \Tr (\frac{1}{2(\bar{x}^-_0)^2}[x'''^-, x'''^p]^2 
+ \frac{1}{4}[x'''^-, x'''^i]^2
-\frac{1}{2(\bar{x}^-_0)^2}[x'''^p, x'''^q]^2
-\frac{1}{2}[x'''^p, x'''^i]^2
-\frac{(\bar{x}^-_0)^2}{8}[x'''^i, x'''^j]^2 \n
&& \qquad \qquad -\frac{i}{2\sqrt{2}} \bar{\psi}''' \Gamma^0 [x'''^-, \psi''']
-\frac{i}{2} \bar{\psi}''' \Gamma^p [x'''_p, \psi''']
-\frac{i}{2\sqrt{2}} \bar{x}^-_0 \bar{\psi}''' \Gamma_{10 i} [x'''^i, \psi'''])
\end{eqnarray} 
in $T \to \infty$ limit.
By redefining
\begin{eqnarray}
&&x'''^i \to \frac{2^{\frac{1}{4}} \sqrt{T}}{\sqrt{\bar{x}^-_0}}x'''^i \n
&&x'''^p \to \frac{\sqrt{\bar{x}^-_0 T}}{2^{\frac{1}{4}}}x'''^p \n
&&x'''^- \to 2^{\frac{1}{4}} \sqrt{\bar{x}^-_0 T}x'''^- \n
&&\psi''' \to \frac{2^{\frac{1}{8}} T^{\frac{3}{4}}}{(\bar{x}^-_0)^{\frac{1}{4}}} \psi''',
\end{eqnarray}
we obtain
\begin{equation}
S=\Tr(\frac{1}{2}[x'''^-, x'''^I]^2 -\frac{1}{4}[x'''^I, x'''^J]^2 
-\frac{i}{2}\bar{\psi'''} \Gamma^0 [x'''^-, \psi'''] 
-\frac{i}{2}\bar{\psi'''} \Gamma^p [x'''_p, \psi''']
-\frac{i}{2} \bar{\psi'''} \Gamma^{10 i} [x'''_i, \psi'''].
\label{preBFSS}
\end{equation}
The background in $x'''^{-}$ is modified, where $\frac{1}{\sqrt{T}}R' \to R'$.
By using the identities (\ref{id}), we can rewrite (\ref{preBFSS}) and obtain the action of BFSS matrix theory with finite $n$,
\begin{equation}
S=\int^{\infty}_{-\infty} d\tau 
\tr( \frac{1}{2}(D_0x^I)^2 -\frac{1}{4}[x^I, x^J]^2
+\frac{1}{2}\bar{\psi} \Gamma^0 D_0 \psi
-\frac{i}{2} \bar{\psi} \Gamma^p [x_p, \psi]
-\frac{i}{2} \bar{\psi} \Gamma^{10 i}[x_i, \psi]).
\end{equation}
We have used $\tilde{R'} = \infty$ because $R' \to 0$ when $T \to \infty$. In DLCQ limit of our model, we see that $X^-$ disappears and $X^+$ changes to $\tau$ as in the case of the light-cone gauge fixing of the membrane theory. 

The way to take DLCQ limit (\ref{StartLimit}) - (\ref{SummaryLimit}) is essentially the same as in the case of the Hermitian model \cite{MModel} because the limit realizes the "novel Higgs mechanism" \cite{Lorentz0}.

\section*{Acknowledgements}
We would like to thank K. Hashimoto, H. Kawai, K. Murakami, F. Sugino, A. Tsuchiya and K. Yoshida for valuable discussions.

\end{document}